\documentclass[aps,prl,twocolumn,superscriptaddress]{revtex4}

\usepackage{amsmath}
\usepackage{amssymb}
\usepackage{graphicx}
\usepackage{slashed}
\usepackage{dsfont}

\begin{document}

\title{$\tau$ decay and the structure of the $a_1$}
\author{M. Wagner}
\affiliation{Institut f\"ur Theoretische Physik, Universit\"at Giessen, Germany}
\author{S. Leupold}
\affiliation{Institut f\"ur Theoretische Physik, Universit\"at Giessen, Germany}
\affiliation{Gesellschaft f\"ur Schwerionenforschung, Darmstadt, Germany}
\date{\today}

\begin{abstract}
We analyse the decay $\tau\rightarrow\pi\pi\pi\nu$ based on the recently developed techniques to generate axial-vector resonances dynamically. Under the assumption that the $a_1$ is a coupled-channel meson-molecule, the spectral function is described surprisingly well by adjusting only one free parameter. Including, in addition, an elementary $a_1$ corrupts the results.
\end{abstract}

%\keywords{}

\maketitle

\begin{figure}
\begin{center}
\begin{tabular}{cc}
\includegraphics[width=3.5cm]{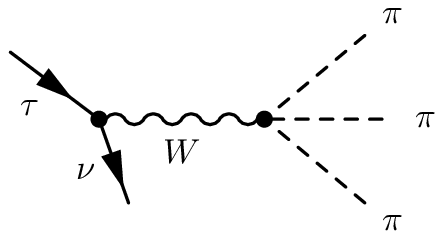} & \includegraphics[width=3.5cm]{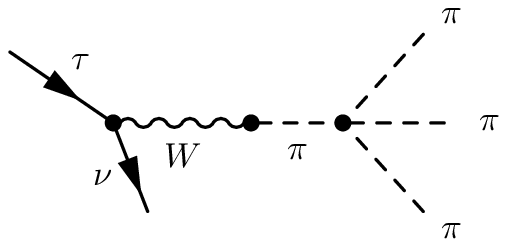} \\
\includegraphics[width=3.5cm]{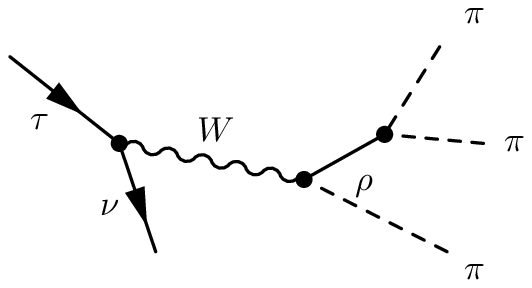} & \includegraphics[width=3.5cm]{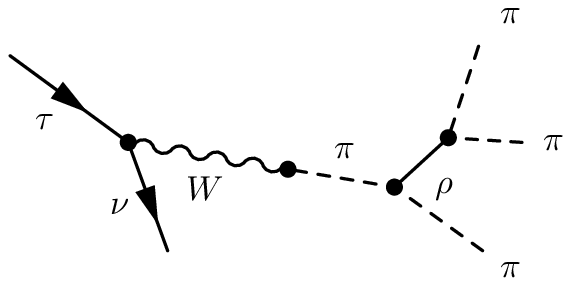} \\
\includegraphics[width=3.5cm]{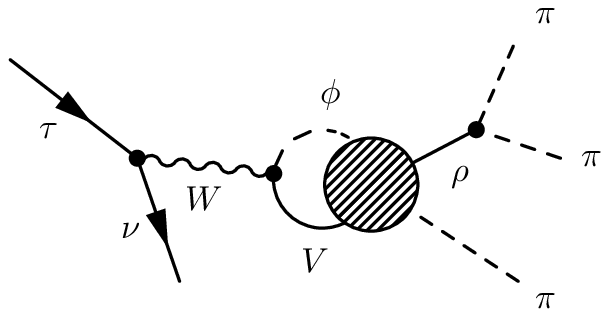} & \includegraphics[width=3.5cm]{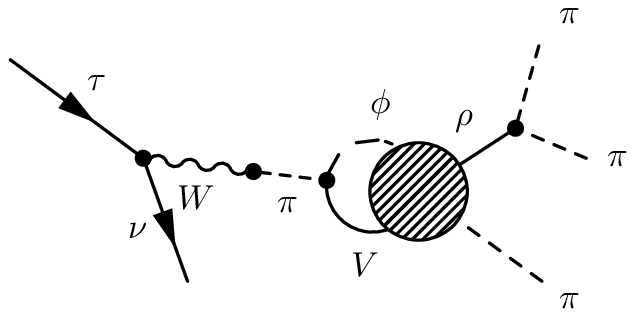} \\
\includegraphics[width=3.5cm]{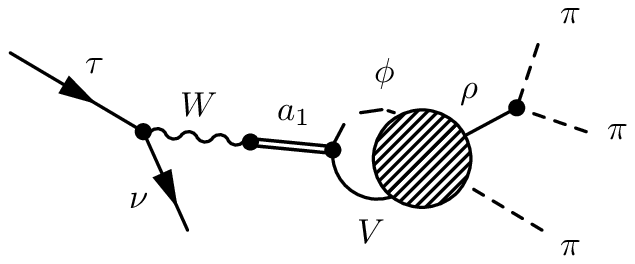} & \includegraphics[width=3.5cm]{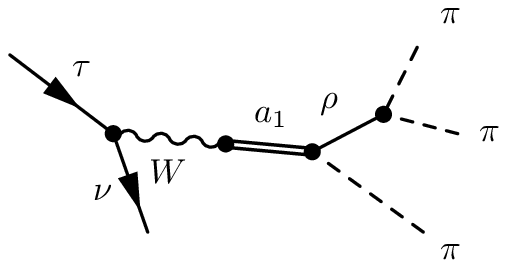} 
\end{tabular}
\caption{Relevant diagrams for the decay $\tau^- \rightarrow \pi^-\pi^0 \pi^0\nu$. The lower two diagrams contribute only if the $a_1$ is included explicitly. $\phi$ and $V$ correspond to intermediate Goldstone bosons and vector meson ($\pi\rho$ or $KK^\ast$). The blob represents the solution of the Bethe-Salpeter equation with the respective kernel. For the WT term alone this leads to the iteration of loop diagrams shown in Fig. \ref{bspic}.}\label{diags}
\end{center}
\end{figure}

\begin{figure}
\begin{center}
\includegraphics[width=8cm]{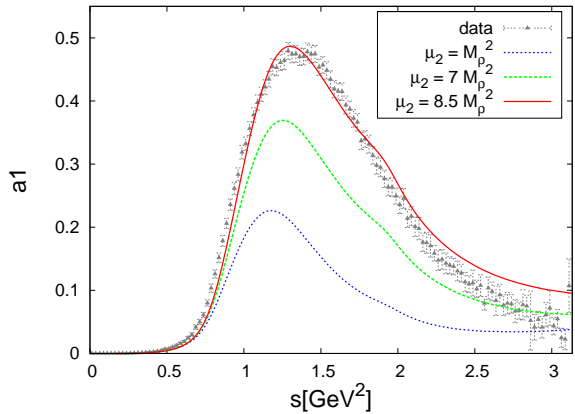}
\caption{Spectral function for the decay $\tau\rightarrow 2\pi^0 \pi$ calculated by using different a subtraction point for the first loop ($\mu_2$) while keeping $\mu_1=M_\rho^2$ in the scattering amplitude. Data are taken from \cite{aleph1}.}\label{subp}
\end{center}
\end{figure}

One aim of elementary particle physics is to get insight into the nature of the hadronic resonances. The constituent quark model has been very successful in describing part of the observed hadron spectrum, especially for heavy-quark systems. On the other hand, especially in the light-quark sector, there is still a lively debate about the nature of many hadronic states.\\
In this letter we present strong indications that the axial-vector meson $a_1(1260)$ \cite{Yao:2006px} is a coupled-channel meson-molecule and not a quark-antiquark state. There are essentially three key ingredients which form the basis of this strong statement:\\
1. The $a_1$ is seen in the decay $\tau\rightarrow 3\pi\nu_{\tau}$. This process is free of hadronic initial state interactions which typically makes the extraction of resonance parameters complicated and often model dependent. It is important to stress that there exist excellent data for this $\tau$ decay process \cite{aleph1}.\\
2. In the proposed molecule scenario the weak current couples (dominantly - see below) to two-particle states, namely $\pi\rho$ and $K^\ast K$. These states are subject to final state interactions which form the resonance seen in the $3\pi$ data. Following \cite{lutz2} (see also \cite{osetaxial}), we describe this final state interaction by a Bethe-Salpeter equation with the kernel fixed by the lowest order interaction of a chiral expansion, the Weinberg-Tomozawa (WT) interaction \cite{wt1,wt2}
\begin{equation}
\mathcal L_{WT} = -\frac{1}{16F_0^2}\text{Tr}[[V^\mu,\partial^\nu V_\mu][\phi,\partial_\nu\phi]],
\end{equation}
where $V_\mu$ is the vector meson octet, $\phi$ is the Goldstone boson field and $F_0$ is the pion decay constant in the chiral limit (concerning chiral perturbation theory (CHPT), see e.g. \cite{gl84,gl2,scherer}). This framework has never been applied to the description of the $\tau$ decay.\\
3. In order to make a decisive statement about the nature of the $a_1$, we perform two calculations. The first one is within the already discussed molecule scenario. In a second calculation we include - in addition to the WT term - an elementary $a_1$ field. It is important to perform both calculations to see how decisive the chosen reaction (here $\tau$ decay) actually is. Of course, data of excellent quality are important for such a task. That kind of comparison of the two scenarios with $\tau$ decay data has never been performed before.\\
Before going into the details of the calculations, we want to put our work in the broader context of the existing literature. Unitary extensions of chiral perturbation theory have shed new light on the structure of several resonances. The low lying scalars ($\sigma$,$f_0(980)$,$a_0(980)$,$\kappa(900)$), for example, appear as bound states in such calculations \cite{osetscalar,nod}. Similar works have been done in the meson-baryon sector (see e.g. \cite{Kaiser:1995eg,Kaiser:1995cy,GarciaRecio:2003ks} and references therein), which suggest a number of $J^P=\frac{1}{2}^-$ baryon resonances to be generated dynamically, in particular the $\Lambda(1405)$ and $N^\ast(1535)$. Studying the interaction of the pseudoscalar mesons with the decuplet of baryons \cite{bardyn3,bardyn5} also led to the generation of many known $J^P=\frac{3}{2}^-$ resonances, e.g. the $\Lambda(1520)$. Recent works applied the approach to the interactions of the octet of pseudoscalar mesons with the nonet of vector mesons \cite{lutz2,osetaxial}, as already mentioned. The only free parameter in the calculation enters through the regularisation of the loop integral in the Bethe-Salpeter equation. Poles have been found, which have been attributed to the axial-vector mesons, but no direct comparison to decay and scattering data has been performed. A comparison of the pole position and width is necessarily indirect and depends on the model, which is used to extract these quantities from the actual observable quantities. In addition, the height of the scattering amplitude, or in other words, the strength of the interaction, is not tested in this way. In the following we use this framework to describe the final state interaction of Goldstone boson and vector meson produced in $\tau$ decays.\\
The $a_1$ is especially interesting as it is considered to be the chiral partner of the $\rho$. One expects a chiral partner for every particle from chiral symmetry. Due to the spontaneous symmetry breaking, one does not find degenerate one-particle states with the appropriate quantum numbers. Nevertheless, the chiral partners have to exist, not necessarily as one-particle states, but at least as multi-particle states. Unmasking the $a_1$ as a bound state of a vector meson with a Goldstone boson would therefore approve its role as the chiral partner. In the meson-meson and meson-baryon scattering examples mentioned before, one can also see that some of the dynamically generated resonances would qualify as the chiral partners of the scattered particles, although the question of the chiral partners for these particles is not as clear as for the $a_1$ and the $\rho$. In addition, the $a_1$ is accessible by a process, which is free of hadronic initial state interactions in contrast to, for example, $\sigma$ or $N^\ast(1535)$. Hence, more solid conclusions can be drawn since one has on the one hand side good quality data and, on the other hand side, a framework in which both scenarios ($a_1$ as a molecule vs. $a_1$ as an elementary state) can be explored.\\
The energies involved in the $\tau$ decay are well beyond 1$\,$GeV, and the decay is dominated by resonance structures. Thus, we can not expect pure CHPT to work for the whole energy region, covered by the $\tau$ decay (see e.g. \cite{Colangelo:1996hs}). Including the vector mesons at tree level, one can also not produce an axial resonance. In case we are not including the $a_1$ explicitly, the picture we promote is that the decay is dominated by the decay into Goldstone boson and vector meson. The structure, which is usually attributed to the $a_1$, is generated by the rescattering process of the vector meson and the Goldstone boson. Since there are two channels with the quantum numbers of the $a_1$, namely $\rho\pi$ and $K^\ast K$, one has to solve a coupled channel problem. In principle, there are also final state interactions between the pions, which we assume to be negligible. This assumption is supported by the success of the model and by the observation that at tree level the direct three pion process is much less important than the $\rho\pi$ channel.\\
The processes, we include in the first calculation, correspond to the upper six diagrams in Fig. \ref{diags}.
\begin{figure}
\begin{center}
\includegraphics[width=7cm]{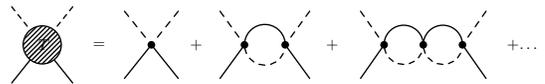}
\caption{Iteration of loop diagrams, corresponding to the approximation to the Bethe-Salpeter equation by using the WT term as kernel.}\label{bspic}
\end{center}
\end{figure}
The first two diagrams are the lowest order contributions from CHPT, the next two correspond to the lowest order diagrams including the vector mesons and the fifth and sixth diagram describe the rescattering process, which is driven by the WT term (see Fig. \ref{bspic}). The couplings describing the decay of the $W$-boson into vector meson and Goldstone boson are related by chiral symmetry breaking to the decay of the $\rho$ into dileptons and two pions, respectively. We refer to \cite{vecrep,tenrep} for further details and in particular for the definition of the coupling constants. In the present work we use $f_V=0.154\,\text{GeV}/M_\rho$ for the coupling of the $\rho$ to a photon and $g_V=0.069\,\text{GeV}/M_\rho$ for the coupling of the $\rho$ to two pions. These numbers are based on the experimental values for the mentioned decays. Thus, all coupling constants are fixed by chiral symmetry breaking and the properties of the $\rho$. We note that in \cite{vecrep} the authors also give a theoretical estimate for $f_V$ and $g_V$, which slightly deviates from the measured values. The dependence of the results on this choice will be discussed in a forthcoming paper \cite{unsers}.\\
Using vector fields as interpolating fields for the vector mesons leads to an unreasonable high-energy behaviour of the spectral function for the $\tau$ decay. Thus, we included higher order corrections in order to cure the high energy behaviour \cite{vecrep,Bijnens:1995ii}, which means that we additionally include $\mathcal O(q^4)$ expressions describing the direct decay into three pions. Here and in the following $q$ is the momentum of the Goldstone bosons in a chiral counting.\\
For this work, we took the scattering amplitude from \cite{lutz2}. We also investigated the dependence of the results on this choice, by taking, for example, the scattering amplitude from \cite{osetaxial}, which will be discussed in \cite{unsers}. Of course, the loops which appear in Fig. \ref{diags} and Fig. \ref{bspic} need renormalisation. In \cite{lutz2} the authors use crossing symmetry arguments in order to fix the subtraction point $\mu_1$ of the loop diagrams (see Fig. \ref{bspic}) at $\mu_1=M_\rho^2$. Looking at the third row of diagrams in Fig. \ref{diags}, one sees that one encounters an additional loop diagram, which is the first loop, containing the decay vertex of the $W$-boson. There is no reason to choose the subtraction point $\mu_2$ for that loop the same as for the loops in the scattering amplitude. Changing $\mu_2$ acts as a higher order correction to the vertex of the reaction $W$-boson to hadrons and is independent of the scattering amplitude. In Fig. \ref{subp} we show the results for different values of $\mu_2$ keeping $\mu_1=M_\rho^2$.
One sees that a peak appears for all values of the only free parameter $\mu_2$. Adjusting $\mu_2$, one influences the height and the width of the peak at the same time, and the choice $\mu_2=8.5M_\rho^2$ reproduces the data quite well. We note that the subtraction point at $8.5M_\rho^2$ approximately corresponds to a cutoff of 1$\,$GeV in a cutoff scheme. \\
In the calculation, we show, we folded the two-particle propagator in the loops with a spectral function (taken from \cite{spec}) for the vector particles. Including the spectral distribution of the vector particles leads to a small broadening of the peak and a small shift to the right. In addition, it smoothens the kink resulting from the threshold effect of the $K^\ast K$ channel \cite{unsers}.\\
Neglecting the coupled channel structure and considering only the $\rho\pi$ channel does not influence the results much. In this case some strength is missing at higher energies ($s\gtrsim 1.4 \,\text{GeV}^2$), but the peak definitely remains \cite{unsers}.\\
In a second calculation we explicitly introduced the $a_1$ in our calculation. In contrast to \cite{pich}, where the width of the $a_1$ is parametrised, we generated the width by the decay of the $a_1$ into Goldstone bosons and vector mesons. In addition, we still include the WT term, since there is no reason to neglect it. A similar calculation without including the WT term has been done in \cite{linsig} in the framework of the linear sigma model. We note that including the WT term and the elementary $a_1$ is not double counting, since integrating out the $a_1$ would lead to a term of at least $\mathcal O(q^2)$, whereas the WT term is $\mathcal O(q^1)$. The processes we include in the second calculation are all diagrams shown in Fig. \ref{diags}. The scattering amplitude of the final state, represented by the blob, is different to the calculation before, since the kernel also contains the $a_1$ interaction (see Fig. \ref{a1kern}).
\begin{figure}
\begin{center}
\includegraphics[width=6cm]{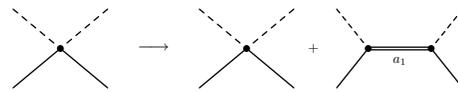}
\caption{The left hand side shows the WT term alone, whereas the right hand side shows in addition the s-channel diagram for an explicit $a_1$. In case the $a_1$ is included explicitly, we replace each dot in Fig. \ref{bspic} by the sum of the WT term and the s-channel diagram.}\label{a1kern}
\end{center}
\end{figure}
Microscopically such an explicit $a_1$ is a quark-antiquark state. On the hadronic level this intrinsic structure is not resolved and we treat the $a_1$ as an elementary field. The Lagrangian describing the decay of the $a_1$ into Goldstone boson and vector meson is given by
\begin{equation}\label{eq1}
\mathcal L_{AV\phi} = i c_1\text{Tr}[V^{\mu\nu}[A_\mu ,u_\nu]] + i c_2\text{Tr}[A^{\mu\nu}[V_\mu, u_\nu]]\,,
\end{equation}
where $V_\mu(A_\mu)$ is the vector (axial-vector)-meson nonet, $V_{\mu\nu}(A_{\mu\nu})$ is the field strength of the vector (axial-vector) mesons and $u_\mu= iu^\dagger D_\mu U u^\dagger$ with the usual non-linear representation of the pseudoscalar mesons $U$ and the covariant derivative $D_\mu$ \cite{scherer}. For $c_1 = -\frac{1}{4}$ and $c_2=-\frac{1}{8}$ Eq.(\ref{eq1}) leads to the same expressions which have been found in \cite{oseta1,meissnera1}.
\begin{figure}
\begin{center}
\includegraphics[width=8cm]{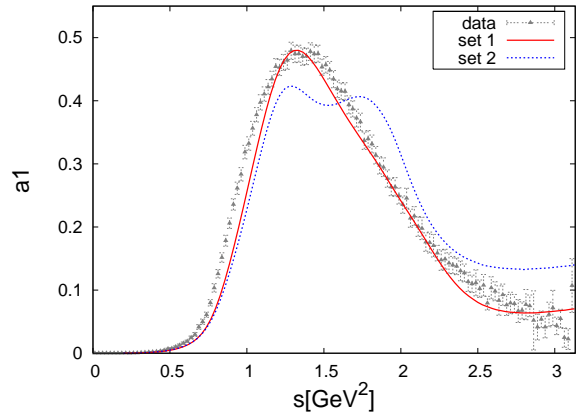}
\caption{Spectral function for the decay $\tau\rightarrow 2\pi^0 \pi$ including the $a_1$ with different sets of parameters in comparison to data from \cite{aleph1}.}\label{a1}
\end{center}
\end{figure}
\begin{table}
\begin{center}
\begin{tabular}{|c||c|c|c|c|c|c|}
\hline
 & $M_{a_1}$[GeV] &  $f_A$ &  $c_1$ &  $c_2$ & $\mu_1\,$[GeV$^2$] & $\mu_2\,$[GeV$^2$] \\ \hline \hline
set 1 & 1.23  & $\frac{F_0}{\sqrt 2 M_\rho}$  & $-\frac{1}{4}\frac{1}{1.65}$ & $-\frac{1}{8}\frac{1}{1.6}$ & $2M_\rho^2$ & $1.05 M_\rho^2$ \\ \hline
set 2 & 1.21 & $\frac{1.45 F_0}{\sqrt 2 M_\rho}$ & $-\frac{1}{4}\frac{1}{2.4}$ & $-\frac{1}{8}\frac{1}{1.6}$ & $M_\rho^2$ & $2.5M_\rho^2$   \\ \hline
\end{tabular}
\end{center}
\caption{Different sets of parameters for the calculations with explicit $a_1$. $f_A$ describes the coupling of the $a_1$ to the $W$-boson \cite{vecrep}.}\label{a1paras}
\end{table}
We know already from the calculation before that a very small coupling of the $a_1$ and the parameters from the previous scenario would result in a good description. For an explicit $a_1$, however, we expect the size of the coupling to be around the values proposed in \cite{oseta1,meissnera1}. Therefore, we keep the contribution of the WT term small in order to be able to arrive at non-zero coupling constants $c_1$ and $c_2$. Most choices of the parameters lead to a double hump structure and only by finetuning one can merge the two bumps. These two possibilities are displayed in Fig. \ref{a1}. The parameters for these curves are shown in Tab. \ref{a1paras}. The merged bump can more or less describe the data, although the agreement is less satisfying due to the deviations on the left hand side of the bump, i.e. for $s\leq 1.1\,$GeV$^2$. However, we do not want to dwell upon this deviation to the data, but we want to emphasise the strong influence of the WT term. We note that by choosing the subtraction points according to Tab. \ref{a1paras}, we already kept the contribution of the WT small (cf. Fig. \ref{subp}). In order to generate a reasonable width of the $a_1$, the driving term should be the decay into vector meson and Goldstone boson. Since the WT term creates a peak by itself, that description leads to peculiar properties, which can be seen in the appearance of a second peak. Merging two bumps by finetuning the parameters does not seem to be a natural way of describing the data. In other words: Why should an elementary state (i.e. a quark-antiquark state) appear right at the mass where an attractive potential (the WT interaction) has already created a resonant structure in the $\pi\rho$ interaction? This would be very artificial. Of course, these considerations would be invalid, if one could arbitrarily tune the strength of the attractive potential. Its size, however, is model independently fixed by chiral symmetry breaking \cite{wt1,wt2,scherer} \\
Finally, we want to remark that in a scenario without explicit $a_1$ it is possible to systematically improve the calculations by including higher order corrections to the WT term, which also reduces the dependence on the subtraction point $\mu_1$ \cite{unsers}. In addition, these terms are sensible to the Dalitz projection data from \cite{dalitz}, which will also be addressed in \cite{unsers}.\\
To summarise, one finds that without the explicit $a_1$ one has a well behaved model, which describes the data very well and which can be systematically improved. All parameters, except of the subtraction points, are fixed by chiral symmetry breaking and the well known properties of the $\rho$. Including an explicit $a_1$ leads to peculiar properties, if one generates the width by the decay into vector meson and Goldstone boson and by including the WT term. These indications point towards a dynamical nature of the $a_1$ as a coupled-channel meson-molecule.

\begin{acknowledgments}
We would like to thank M. F. M. Lutz and E. Kolomeitsev for useful discussions. We would also like to acknowledge stimulating discussions with E. Oset. We also thank Hasko Stenzel for clarifying our questions concerning the data and we are grateful to U. Mosel for stimulating discussions and continuous support.
\end{acknowledgments}

\bibliography{litfile}

\end{document}